\documentclass[apj,iop]{emulateapj}
\usepackage{apjfonts}
\usepackage{graphicx}
\usepackage{amsmath,amsthm,amsfonts,amssymb}
\usepackage[breaklinks,colorlinks,urlcolor=magenta,citecolor=blue,linkcolor=blue]{hyperref}
\usepackage{natbib}

\shorttitle{Dielectronic recombination satellite lines}
\shortauthors{Shah et al.}

\begin{document}

\title{Polarization of K-shell dielectronic recombination satellite lines of F\lowercase{e} \textsc{XIX--XXV} and its application for diagnostics of anisotropies of hot plasmas}

\author{Chintan Shah\altaffilmark{1, 2, $\ast$}, Pedro Amaro\altaffilmark{1, $\dagger$}, Ren\'e Steinbr\"{u}gge\altaffilmark{2, $\S$}, Sven Bernitt\altaffilmark{2, 3}, Jos\'e R. Crespo L\'opez-Urrutia\altaffilmark{2}, and Stanislav Tashenov\altaffilmark{1}}

\affil{$^1$Physikalisches Institut der Universit\"at Heidelberg, 69120 Heidelberg, Germany}
\affil{$^2$Max--Planck--Institut f\"{u}r Kernphysik, 69117 Heidelberg, Germany}
\affil{$^3$Institut f\"ur Optik und Quantenelektronik, Friedrich-Schiller-Universit\"at, 07743 Jena, Germany}

\altaffiltext{$\ast$}{\href{mailto:chintan@mpi-hd.mpg.de}{chintan@mpi-hd.mpg.de}}
\altaffiltext{$\dagger$}{Current address: LIBPhys-UNL, FCT-UNL, P-2829-516, Caparica, Portugal}
\altaffiltext{$\S$}{Current address: Deutsches Elektronen-Synchrotron (DESY), 22607 Hamburg, Germany}


\begin{abstract}
We present a systematic measurement of the X-ray emission asymmetries in the K-shell dielectronic, trielectronic, and quadruelectronic recombination of free electrons into highly charged ions. Iron ions in He-like through O-like charge states were produced in an electron beam ion trap, and the electron-ion collision energy was scanned over the recombination resonances. Two identical X-ray detectors mounted head-on and side-on with respect to the electron beam propagation recorded X-rays emitted in the decay of resonantly populated states. The degrees of linear polarization of X-rays inferred from observed emission asymmetries benchmark distorted-wave predictions of the Flexible Atomic Code (FAC) for several dielectronic recombination satellite lines. The present method also demonstrates its applicability for diagnostics of energy and direction of electron beams inside hot anisotropic plasmas. Both experimental and theoretical data can be used for modeling of hot astrophysical and fusion plasmas. 
\end{abstract}



\section{Introduction}\label{sec:intro}
 
The major part of all visible matter in the Universe is in a state of highly-ionized hot plasma. 
In most astrophysical plasmas the electrons reach thermodynamic equilibrium where their velocities are distributed isotropically following the Maxwell-Boltzmann law. 
Such plasmas emit isotropic and unpolarized radiation. 
However, at very high electron energies, the electron velocity distribution may become anisotropic. 
Anisotropic plasmas are found prominently in solar flares~\citep{haug1972,haug1979,haug1981,emslie2008}. X-rays emitted from such plasmas are usually highly anisotropic and polarized~\citep{inal1987,inal1989}. 
Anisotropic plasmas are also found in pulsars~\citep{weisskopf1976,weisskopf1978,vadawale2017}, in neutron stars~\citep{weisskopf2006}, and around accreting black holes~\citep{dovciak2004,dovciak2008,nayakshin2007} (see also~\citet{kallman2004} and references therein).  
They also appear prominently in laboratory plasmas where electron-ion collisions are directional, e.g., tokamak plasmas~\citep{fujimoto1996}, ECR plasmas~\citep{iwamae2005}, and laser-produced plasmas~\citep{kieffer1993}.  
Since the X-ray polarization is caused by the plasma anisotropy, polarization measurements can provide valuable and often unique insights into physical conditions of such plasmas. 
It can reveal the distributions of nonthermal or suprathermal electrons, an orientation of electron or ion beams in plasmas, and it can likewise tell us orientations of magnetic fields present inside the plasmas. 
Indeed, in astrophysics, X-ray polarization measurements are the only way to derive information on the geometry of angularly unresolved sources~\citep{krawczynski2011}.

Interpretation of the spectral features from hot anisotropic plasmas requires understanding of different atomic excitation mechanisms that can lead to polarized X-ray emission. 
Polarized X-ray continuum can be produced by bremsstrahlung and radiative recombination~\citep{tashenov2006,tashenov2011,tashenov2014}; at the same time, direct electron-impact excitation, radiative cascade, and dielectronic recombination can produce characteristic polarized X-ray lines~\citep{beiersdorfer1996,takacs1996,beiersdorfer1997,nakamura2001,shlyaptseva1998,shlyaptseva1999}. 
The construction of a reliable model requires a systematic understanding of each of these components, and calculations of them have to be benchmarked against experiments. 
The aim of this paper is to study the polarization of Fe K$\alpha$ X-ray lines produced by dielectronic recombination and to provide experimental benchmarks under well-controlled conditions.

Dielectronic recombination (DR)~\citep{burgess1964} is the dominant electron-ion collision process, active in both photoionized and collisionally ionized high-temperature plasmas. 
As a line formation mechanism in plasmas, the total cross sections of DR surpass those of all other processes by orders of magnitude. 
DR is a resonant two-step process, in which a free electron is captured into an ion while a bound electron is simultaneously excited. Thereby, an intermediate excited (autoionizing) state, or resonance, is formed. DR process is completed by radiative emission, which is the dominant channel in highly charged ions, as opposed to Auger emission. 
It strongly influences total recombination rates, and thus the charge balance of hot plasmas~\citep{dubau1980}. 
DR also provides resolved spectral lines which can be utilized for plasma temperature and density diagnostics~\citep{widmann1995,kato1998,porquet2010}. 
Therefore, an exact knowledge of DR cross sections is needed not just to understand astrophysical observations (e.g., recent results of~\citet{hitomi2016}), but also to benchmark widely used X-ray spectral models such as AtomDB~\citep{foster2012}, SPEX~\citep{kaastra1996}, and CHIANTI~\citep{delzanna2015} for hot collisional plasmas.

Beyond DR, more exotic higher-order resonant recombinations such as trielectronic recombination (TR) or quadruelectronic recombination (QR) are also relevant. 
Here, a minimum two or even three bound electrons can be simultaneously excited upon the capture of a free electron, leading to TR and QR, respectively, see Fig.~\ref{fig:2Dspec}.
The recombination rates in low-temperature photoionized plasmas were shown to be dominated by intra-shell ($n$=2--2) TR~\citep{schnell2003,orban2010}. 
Similarly, inter-shell ($n$=2--1) TR was recently measured and found to have a sizable and even greater strength compared to DR for astrophysically relevant low-Z ions~\citep{beilmann2011,beilmann2013}. 
It is therefore also necessary to include these channels in the plasma model to predict accurately ionization balance~\citep{beiersdorfer2015,shah2016}.

Several experimental and theoretical investigations were performed on DR of highly charged ions. 
The intra-shell low-energy DR rate coefficients were measured at storage rings~\citep{savin2002,savin2003,savin2006,savin2007,orban2010}, while inter-shell high-energy DR strengths and rate coefficients were measured at electron beam ion traps (EBITs)~\citep{beiersdorfer1992,fuchs1998,smith2000,yao2010,ali2011,hu2013}. 
However, due to the directionality in electron-ion collisions as well as due to the geometry of these devices, emitted X-rays are usually anisotropic and polarized. 
In such cases, accurate values of the line polarization are important for interpreting high-resolution measurements and extracting accurate recombination rate coefficients.

Apart from the application relevance of DR in plasmas, the interest in DR also arises from the point of view of understanding electron-electron interaction in the strong Coulomb fields of highly charged ions. 
The relativistic effect in the electron-electron interaction is known as the Breit interaction~\citep{breit1929}. It includes magnetic interactions and the retardation in the exchange of a virtual photon between the electrons. 
{While its effect on energy levels is considered as a modest correction, it significantly changes the DR cross sections and the polarization of DR X-rays for high-Z ions~\citep{nakamura2008,fritzsche2009,hu2014,shah2015,nakamura2016,amaro2017}. }

Here, we present a comprehensive experimental and theoretical study on the polarization of X-rays produced by {KLL} DR of highly charged Fe ions. 
In KLL DR, a bound K-shell electron is excited to L-shell by recombination of a free electron from continuum to L-shell. 
An entire set of observed recombination resonances of Fe~\textsc{XIX--XXV} ions are analyzed to systematically benchmark distorted-wave predictions of Flexible Atomic Code (FAC:~\citet{gu2008}) on line energies, resonance strengths, and polarization of DR X-rays.  
In the course of this work, we also demonstrate an application of the present data for diagnostics of anisotropies of hot plasmas.

The paper is organized as follows. 
The experimental arrangement is described in~Sec.~\ref{sec:exp}. 
In~Sec.~\ref{sec:data}, the data analysis method and results are presented. 
A theoretical outline with a short account of basic formulas is given in~Sec.~\ref{sec:theory} describing the calculation of DR strengths and polarization of DR X-rays. There we also outline the comparison between the experimental and theoretical results.  
In~Sec.~\ref{sec:application}, we show an example of the application of measured polarization data for diagnostics of hot anisotropic plasmas, namely we diagnose the electron cyclotron motion in an EBIT. 
A summary and conclusion of the present work are given in~Sec.~\ref{sec:summary}.
All theoretical values are presented in appendix~\ref{appendix} which is also provided as machine-readable tables.

\section{Experiment}\label{sec:exp}

%
\begin{figure*}
	\centering
	\includegraphics[width=0.92\textwidth]{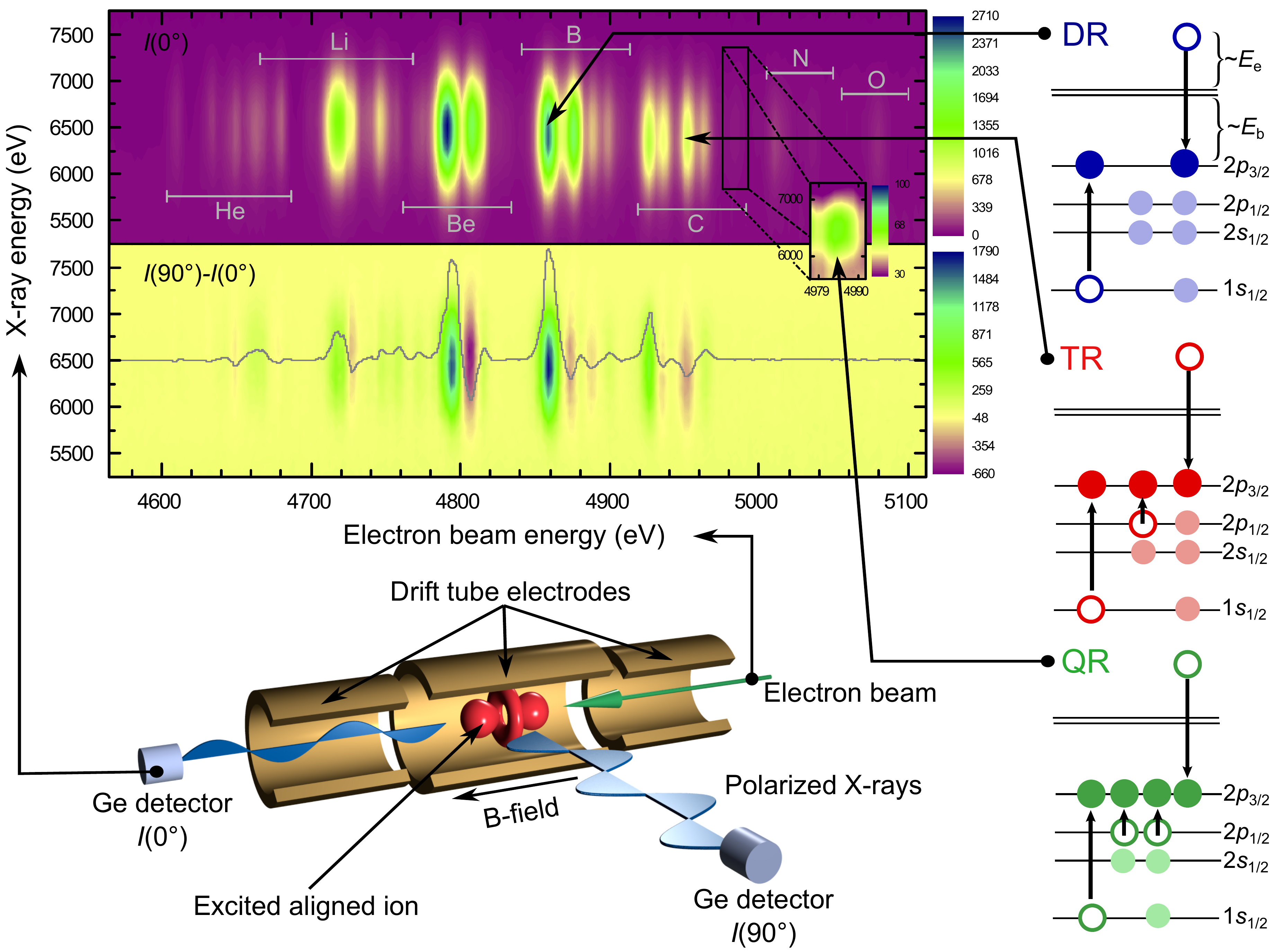}
	\caption{
		Experimental setup and data map of Fe~\textsc{XIX-XXV} ions. 
		A compressed monoenergetic electron beam of an EBIT produces and traps highly charged Fe ions. 
		The electron-ion collision energy $E_\mathrm{e}$ was tuned into the {KLL} recombination resonances and the subsequent X-ray fluorescences were observed by two germanium detectors at $0^\circ$ and $90^\circ$ with respect to the electron beam propagation axis. 
		Upper plot: Intensity of X-rays observed along the beam axis as a function of the electron beam and X-ray energies. 
		Lower plot: A difference in the X-ray intensities observed by the two detectors, and the solid gray line shows this difference qualitatively. 
		Some examples of individual resonances due to DR, TR, and QR recombination channels are also marked and they are identified by the charge state of the initial ion. (Modified from~\citet{shah2016})
	} 
	\label{fig:2Dspec}
\end{figure*}
\begin{table}[b]
	\centering
	\caption{Operational parameters of the FLASH-EBIT in the present experiment. (a) and (b) corresponds to two different measurements.}
	\begin{tabular}{ll}
		\hline \hline
		Parameter & Value \\
		\hline
		Electron beam current& (a) {100}~{mA}  (b) {70}~{mA} \\
		Magnetic field& {6} {T} \\
		Trap offset potential & (a) {110}~{V} (b) {20}~{V}\\
		Scan energy & {4.5} -- {5.2}~{keV} \\ 
		Sweep rate & {1.16}~{eV/s}\\
		Beam radius& $\approx$ {24.4} {$\micron$}~\citep{herrmann1958}\\
		& 22.6 $\pm$ 4.2 {$\micron$} (See Sec.~\ref{sec:application})\\
		Trap length& {50}~{mm} \\
		Dump cycle&(a) {20}~{s}  (b) {5}~{s}\\
		Pressure of gas injector & (a) {$1.2 \times 10^{-6}$}~{mbar}\\
		& (b) {$1.1 \times 10^{-7}$}~{mbar}\\
		\hline \hline
	\end{tabular}
	\label{tab:pars_fe}
\end{table}

The experiment was performed at an electron beam ion trap (FLASH-EBIT) at the Max-Planck-Institut f\"{u}r Kernphysik, Heidelberg, Germany. The FLASH-EBIT, described in detail by~\citep{epp2010}, was designed for photonic studies of highly charged ions and has been used in several experiments at synchrotrons and free electron lasers~\citep{epp2007, bernitt2012, rudolph2013, steinbrugge2015}. FLASH-EBIT uses a nearly unidirectional monoenergetic electron beam to produce highly charged ions through successive electron impact ionization. The negative space charge potential of the electron beam traps ions radially; while along the electron beam direction ions are trapped axially by the different potentials applied to a set of surrounding drift tubes. Further potentials are applied to these drift tubes to change the electron beam energy without affecting the trap potentials. A high magnetic field of {6} {T} at the trap compresses the electron beam to a radius of $\approx$~{24}~$\micron$. This value is calculated according to~\citep{herrmann1958}, and is confirmed within the present experiment (see Sec.~\ref{sec:application}). Iron was continuously injected into the trap in the form of iron pentacarbonyl molecules with a help of the differential pumping system. The pressure at the injection and trap was $1.1 \times 10^{-7}$ mbar and $1.5 \times 10^{-10}$ mbar, respectively. Overall, FLASH-EBIT was operated under ultrahigh vacuum conditions ($\leq 5 \times 10^{-10}$ mbar).

For steady-state conditions~\citep{penetrante1991}, the electron-ion collision energy was swept over the range of DR, TR, and QR resonances exciting {K}-shell electrons in iron ions. The electron beam energy was continuously scanned from  4.5~keV to 5.2~keV at a sweep rate of 1.16~eV/s in a triangular shape waveform. The experimental parameters were optimized for two criteria as follows:

\noindent(a) A deeper trap with a higher density of ions in high charge states (e.g., He-, Li-like ions): For this purpose, the deeper axial trap was used at {100}~{mA} electron beam current. The trap was dumped repeatedly for one second every {20}~seconds.  With this high beam current and deep trap setting, an electron-ion collision energy resolution of 11~eV FWHM at 5~keV was achieved.

\noindent(b) For observation of the weak and blended higher-order resonances, a higher energy resolution is required. Therefore, in this case, we improved the energy resolution by application of an evaporative cooling technique~\citep{penetrante1991a} in combination with a moderate electron beam current ({70} {mA} here) sufficient for an efficient ionization and recombination yield. This was achieved by lowering the axial potential well applied to the drift tubes. The ion dump cycle was shortened to 5~sec to reduce the accumulation of unwanted heavy ionic species like tungsten and barium emitted from the hot cathode of the electron gun. Additionally, this helped to shift the ionization balance towards lower charge states where TR and QR are more prominent. With this method, an excellent electron-ion collision energy resolution of 6~eV FWHM at 5~keV was achieved. We note that the resolution in this energy range was better than in any previously reported measurements and allowed to distinguish well-resolved TR and QR along with DR resonances. A list of operational parameters of the FLASH-EBIT is summarized in Table~\ref{tab:pars_fe}.

\begin{figure*}
	\centering
	\includegraphics[width=0.92\textwidth]{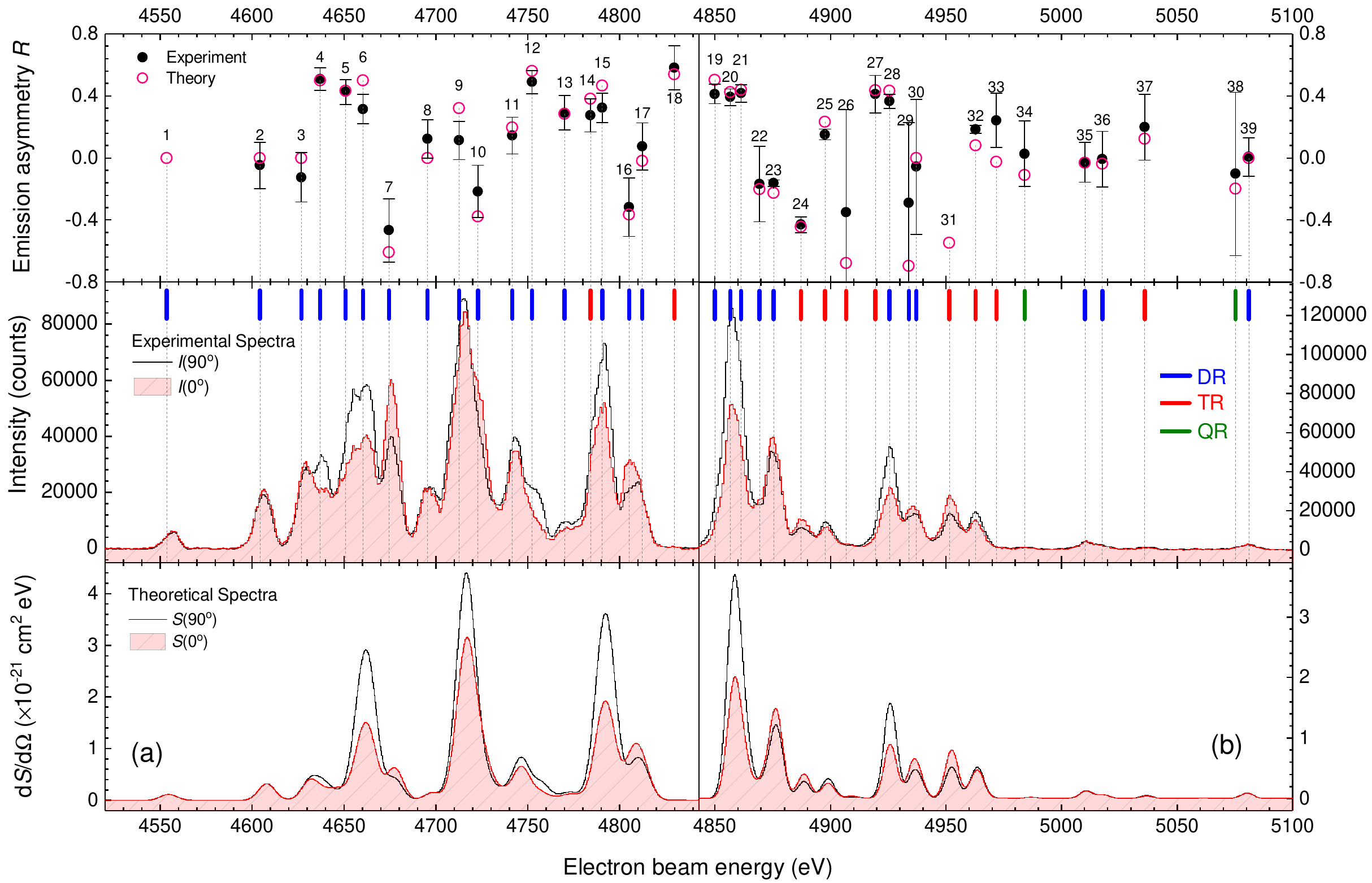}
	\caption{
		{Top panel} shows the extracted emission asymmetries $\mathcal{R}$ (black solid squares) for individual resonances and the corresponding predictions with Breit interaction included (magenta open circles).
		{Middle panel} shows the X-ray intensities registered by the two germanium detector plotted as a function of electron beam energy. Each line is identified by one or more excited state formed by a given charge state and process in Table~\ref{tab:results}. 
		Vertical markers represent the order of recombination. Blue: DR, Red: TR and Green: QR. 
		The background due to radiative recombination was subtracted from both spectra.
		{Bottom panel} shows synthetic spectra -- a differential resonance strength ($dS/d\Omega$) as a function of electron beam energy. 
		The black solid lines and the red shaded areas correspond respectively to the X-rays observed along and perpendicular the electron beam in both middle and bottom panels. 
		The labels (a) and (b) correspond to the two different measurement schemes, their parameters are shown in Tab.~\ref{tab:pars_fe}.
	} 
	\label{fig:1Dspec}
\end{figure*}

To observe the X-ray emission asymmetry, the radiative decay of resonantly exited states has been simultaneously observed by two identical liquid nitrogen cooled solid-state germanium detectors aligned parallel ($0^\circ$) and perpendicular ($90^\circ$) to the electron beam propagation axis, see Fig.~\ref{fig:2Dspec}. Both detectors have an intrinsic X-ray energy resolution of approximately {550}~{eV} FWHM at {13}~{keV}. X-ray detectors at EBITs are typically mounted perpendicular to the electron beam propagation axis. For this experiment, a beryllium window was installed behind the electron beam collector in order to place there another identical germanium detector along the electron beam axis. However, at this observation point, solid angle and X-ray flux are both significantly reduced as compared to the detector mounted perpendicular to the electron beam propagation axis. The solid angle of detection was $8.7\times10^{-3}$~{sr} for the detector at $90^\circ$, whereas the solid angle for X-ray detection along the electron beam was $7.3\times10^{-4}$~{sr}. Hence, the detector at $0^{\circ}$ covered less solid angle compared to the one at $90^{\circ}$ by a factor of $\approx$ 11.9. 

The intensity of X-rays observed at $0^{\circ}$ with respect to electron beam propagation direction is shown in the upper data inset of Fig.~\ref{fig:2Dspec} as a function of the electron beam and the X-ray energies. 
A similar histogram was obtained for the detector at $90^{\circ}$. The X-ray energy was calibrated using the X-ray radiation from an Americium-241 radioactive isotope, whereas the electron beam energy was calibrated using theoretical resonance energies. The visible bright spots are {KLL} DR resonances in He-like to O-like iron ions. Each resonance is usually identified by the initial charge state of iron in which resonant recombination occurred. X-rays due to radiative recombination (RR) into the L-shell ($n$=2) appear as a background. The RR X-ray energy is equal to the sum of the electron energy $E_\mathrm{e}$ and the binding energy $E_\mathrm{b}$ of the L-shell, see Fig.~\ref{fig:2Dspec}.
%

\section{Data Analysis and Results}\label{sec:data}
%
The X-ray intensity as a function of electron beam energy was obtained by integrating the X-rays observed within the X-ray energy interval of the KLL resonances. 
In this way, X-rays emitted in bremsstrahlung and RR into K, M and higher shells were excluded. 
This procedure is essential to reduce the background and observe weak resonances. 
Several resonantly excited X-ray transitions due to DR, TR, and QR were observed in the region of interest, see Fig.~\ref{fig:1Dspec}. They were identified by extensive and detailed calculations using FAC~\citep{gu2008}. The computational details are described later in Sec.~\ref{sec:theory}. 
Corresponding theoretical resonance strengths were also calculated in order to quantify the relative weights that unresolved excited states contribute to the X-ray lines. Figure~\ref{fig:1Dspec} shows several well-resolved DR, TR, and QR satellite transitions.

The electron beam energy can be described by the sum of all accelerating potentials that are applied minus the net space charge potential due to the negatively charged electrons of the beam as well as the positively charged ions accumulated in the trap volume. The energy-dependent part of the space charge potential was calculated at maximum and minimum acceleration potentials and can be considered to be constant along the energy scan~\citep{penetrante1991}. Thus, the electron beam energy scale can be calibrated using theoretical resonance energies. The electron beam energy scale of measurement (a) in Fig.~\ref{fig:1Dspec} was calibrated using X-ray lines 1 and 8, corresponding to energies of {4554.4} {eV} and {4697.8} {eV}, respectively. Similarly, for measurement (b), the X-ray lines 20 ({4858.9} {eV}) and 31 ({4952.2} {eV}) were employed for the calibration.

The observed spectra need to be corrected for the detector solid angles and the background arising from the L-shell RR. 
We consider the intensity of RR as a linear function of the electron beam energy, which we fit and subtract from the total intensity. 
The solid-angle factor can be corrected using a DR transition theoretically known to be isotropic. 
For the measurement (a), we selected the DR resonance $[1s 2s^2]_{1/2}$ (line 1) of initially He-like ions to obtain $\Omega_{0^{\circ}}/\Omega_{90^{\circ}}=0.09\pm0.02$. 
In the measurement (b), the observed isotropic transitions have very low intensities. 
Hence, we calculated the detector solid angles using line 31 -- the TR resonance $[1s2s^22p_{1/2}2p_{3/2}^3]_{J=3/2}$ of initially C-like ions. It has a theoretical intensity ratio of $I_{0^{\circ}}/I_{90^{\circ}}=1.55$. 
We selected it for its higher resonance strength and null influence of the Breit interaction on its angular distribution~\citep{gail1998}. Moreover, our previous experiments on Kr have confirmed the theoretical ratio for this line~\citep{shah2016}. 
The measured ratio of solid angles was found to be $\Omega_{0^{\circ}}/\Omega_{90^{\circ}}=0.082\pm0.003$. 
In both cases, used normalization resonances are well separated from the neighboring transitions. 
After correction for these effects, the corresponding intensity differences indicate an anisotropic X-ray emission for most of the resonances. 
The red shaded area in Fig.~\ref{fig:1Dspec} represents the X-ray intensity along the direction of the electron beam multiplied by the solid-angle factor.

Since the electron beam energy spread follows a normal distribution and the natural width of the excited state is much lower than this energy spread, spectral resonances can be well described by a Gaussian function. 
The amplitude and energies $E_{\mathrm{res}}$ of the resonance lines, including partially blended resonances, were obtained by fitting the measured spectrum with the Gaussian functions. 
Since the fine-structure splitting scales with Z$^4$, resonances in low-Z elements, such as iron, are not well resolved. 
Therefore, the selected energy region contains a high density of excited states. 
For this reason, we used the following criterion to avoid unwanted spurious effects due to possible interference between two or more excited states in the fitting process. 
We considered only those excited states having energy differences larger than half of the collision energy resolution -- namely for part (a) 5 eV and for part (b) 3 eV. 
Moreover, the excited states having relative resonance strength smaller than $\approx$ 5\% compares to the neighboring lines were not considered in the fitting procedure.

By applying the above criteria, the amplitudes and centroids corresponding to each resonance were fitted to the spectrum of X-rays observed perpendicular to the electron beam propagation direction. Due to the larger solid angle of this detector, this spectrum had significantly higher statistics. In the first step of the fitting procedure, the line positions and amplitudes were set free while the widths were fixed to the measured electron beam energy resolution. In the second step, the similar procedure was performed for the spectrum observed along the electron beam propagation direction. However, in this step, the line centroids were fixed to the ones obtained in the first step. The extracted intensities were corrected for the detector solid-angle factor. We stress that the high statistics collected in this experiment allowed us to reliably extract both the line centroids and the intensities of even blended strong resonance lines.

From Fig.~\ref{fig:1Dspec}, it is apparent that most observed X-ray transitions are anisotropic, as their intensities differ when observed along and perpendicular to the electron beam axis. We quantify the emission asymmetries by the ratios
\begin{equation}
\mathcal{R} = \frac{I(90^{\circ})-I(0^{\circ})}{I(90^{\circ})}.
\label{eq:P90}
\end{equation}
The experimental emission asymmetries $\mathcal{R}$ are extracted for each fitted recombination transition and summarized in Tab.~\ref{tab:results} as well as in the uppermost inset of Fig.~\ref{fig:1Dspec}. Uncertainties in the emission asymmetries are taken as a quadrature sum of the uncertainty of measured intensity and the uncertainty of the detector solid-angle correction. 
Moreover, a finite thermal velocity component of the electron beam has a systematic effect on measured emission asymmetries~\citep{gu1999,beiersdorfer2001}. For each of the resonances, we have estimated its effect and found that the measured emission asymmetries were indeed reduced, but only within the uncertainty limits of our measurement. Thus, no corrections for $\mathcal{R}$ were made in this part of the analysis. However, by making a rigorous statistical analysis on the total dataset of observed resonances we can nonetheless determine the thermal velocity component of the electron beam. We demonstrate this later in Sec.~\ref{sec:application}.

\begin{table*}
	\centering
	\scriptsize
	\setlength{\tabcolsep}{1.7em}
	\caption{
		Measured emission asymmetry $\mathcal{R}$ between perpendicular and parallel observation of X-rays emitted due to DR, TR, and QR. 
		First column labels identify resonances in Fig.~\ref{fig:1Dspec} made of a single or an ensemble of unresolved intermediate excited states.  
		The initial ionic charge state before the recombination and the order of the recombination are given in next two consecutive columns, respectively. Intermediate excited states are given in the \textit{j-j} coupling notations. Round brackets stand for the angular momentum of coupled subshells and subscripts after the square brackets denotes the total angular momentum of the state. 
		Measured resonance energies $E_{\mathrm{res}}$ (in eV) from the fits (except fixed, blended, and calibrated ones) are presented. 
		Extracted emission asymmetries $\mathcal{R}$ are compared with the FAC predictions. 
		Experimental uncertainties are given as 1$\sigma$.
	}
	\begin{tabular}{ccclccccc}
		\hline
		Label	&	Initial ion	&	Process	&	Intermediate excited state &	$E_\mathrm{res}$ (Th.)	&	$E_\mathrm{res}$ (Exp.)			&	$\mathcal{R}$ (Exp.)			&	$\mathcal{R}$ (Th.)	\\
		\hline
		1	&	He	&	DR	&	$[1s 2s^{2}]_{1/2}$	&	4554.4	&	calib.			&	norm.			&	0.00	\\
		2	&	He	&	DR	&	$[(1s 2s)_0 2p_{1/2}]_{1/2}$	&	4607.8	&	4604.2	$\pm$	0.5	&	-0.05	$\pm$	0.15	&	0.00	\\
		3	&	He	&	DR	&	$[(1s 2s)_1 2p_{3/2}]_{1/2}$	&	4631.2	&	4626.8	$\pm$	0.7	&	-0.13	$\pm$	0.16	&	0.00	\\
		4	&	He	&	DR	&	$[(1s 2p_{1/2})_1 2p_{3/2}]_{5/2}$	&	4638.5	&	4636.9	$\pm$	1.8	&	0.51	$\pm$	0.07	&	0.50	\\
		5	&	Li	&	DR	&	$[1s 2s^2 2p_{1/2}]_1$	&	4646.8	&	blend	&	0.43	$\pm$	0.08	&	0.43	\\
		&	He	&	DR	&	$[(1s 2p_{1/2})_1 2p_{3/2}]_{3/2}$	&	4658.3	&			&				&		\\
		6	&	He	&	DR	&	$[(1s 2p_{1/2})_1 2p_{3/2}]_{3/2}$	&	4658.3	&	blend	&	0.31	$\pm$	0.10	&	0.50	\\
		&	He	&	DR	&	$[1s (2p_{3/2}^2)_2]_{5/2}$	&	4663.9	&			&				&		\\
		7	&	Li	&	DR	&	$[(1s 2s 2p_{1/2})_{3/2} 2p_{3/2}]_3$	&	4675.5	&	blend	&	-0.47	$\pm$	0.20	&	-0.61	\\
		&	He	&	DR	&	$[1s (2p_{3/2}^2)_2]_{3/2}$	&	4677.5	&				&				&		\\
		8	&	He	&	DR	&	$[1s (2p_{3/2}^2)_0]_{1/2}$	&	4697.8	&	calib.			&	0.12	$\pm$	0.12	&	0.00	\\
		9	&	Li	&	DR	&	$[((1s 2s)_1 2p_{1/2})_{1/2} 2p_{3/2}]_1$	&	4712.0	&	blend	&	0.11	$\pm$	0.12	&	0.32	\\
		&	Li	&	DR	&	$[((1s 2s)_1 2p_{1/2})_{3/2} 2p_{3/2}]_2$	&	4713.8	&				&				&		\\
		10	&	Li	&	DR	&	$[(1s 2s)_1 (2p_{3/2}^2)_2]_1$	&	4718.8	&	blend &	-0.22	$\pm$	0.17	&	-0.38	\\
		&	Li	&	DR	&	$[(1s 2s)_1 (2p_{3/2}^2)_2]_2$	&	4726.0	&				&				&		\\
		11	&	Li	&	DR	&	$[((1s 2s)_0 2p_{1/2})_{1/2} 2p_{3/2}]_2$	&	4745.7	&	blend	&	0.14	$\pm$	0.12	&	0.20	\\
		&	Li	&	DR	&	$[(1s 2s)_1 (2p_{3/2}^2)_0]_1$	&	4749.7	&				&				&		\\
		12	&	Li	&	DR	&	$[1s 2s^2 2p_{3/2}]_2$	&	4751.0	&	blend	&	0.49	$\pm$	0.07	&	0.56	\\
		&	Be	&	DR	&	$[1s 2s^2 2p_{1/2}^2]_{1/2}$	&	4755.8	&				&				&		\\
		&	Li	&	DR	&	$[(1s 2s)_0 (2p_{3/2}^2)_2]_2$	&	4757.4	&				&				&		\\
		13	&	Li	&	DR	&	$[(1s 2s)_1 (2p_{3/2}^2)_2]_1$	&	4767.5	&	blend	&	0.29	$\pm$	0.11	&	0.28	\\
		&	Be	&	DR	&	$[(1s 2s^2 2p_{1/2})_1 2p_{3/2}]_{5/2}$	&	4770.3	&				&				&		\\
		14	&	Li	&	TR	&	$[(1s 2p_{1/2})_0 (2p_{3/2}^2)_2]_2$	&	4781.8	&	4784.0	$\pm$	1.2	&	0.28	$\pm$	0.11	&	0.38	\\
		15	&	Be	&	DR	&	$[(1s 2s^2 2p_{1/2})_1 2p_{3/2}]_{1/2}$	&	4789.1	&	blend	&	0.32	$\pm$	0.09	&	0.47	\\
		&	Be	&	DR	&	$[(1s 2s^2 2p_{1/2})_1 2p_{3/2}]_{3/2}$	&	4789.4	&				&				&		\\
		&	Be	&	DR	&	$[1s 2s^2 (2p_{3/2}^2)_2]_{5/2}$	&	4794.4	&				&				&		\\
		16	&	Be	&	DR	&	$[1s 2s^2 (2p_{3/2}^2)_2]_{3/2}$	&	4806.9	&	blend	&	-0.32	$\pm$	0.19	&	-0.36	\\
		&	Be	&	DR	&	$[1s 2s^2 (2p_{3/2}^2)_0]_{1/2}$	&	4812.5	&				&				&		\\
		17	&	Be	&	DR	&	$[1s 2s^2 (2p_{3/2}^2)_0]_{1/2}$	&	4812.5	&	blend	&	0.08	$\pm$	0.15	&	-0.02	\\
		&	Li	&	TR	&	$[1s 2p_{3/2}^3]_2$	&	4815.6	&				&				&		\\
		18	&	Be	&	TR	&	$[(1s 2s)_1 2p_{1/2}^2 2p_{3/2}]_{3/2}$ 	&	4829.0	&	fix.			&	0.58	$\pm$	0.14	&	0.54	\\
		19	&	B	&	DR	&	$[1s 2s^2 2p_{1/2}^2 2p_{3/2}]_1$	&	4856.9	&	4849.9	$\pm$	0.2	&	0.41	$\pm$	0.06	&	0.50	\\
		20	&	B	&	DR	&	$[(1s 2s^2 2p_{1/2})_0 (2p_{3/2}^2)_2]_2$	&	4858.9	&	calib.			&	0.39	$\pm$	0.05	&	0.42	\\
		21	&	B	&	DR	&	$[(1s 2s^2 2p_{1/2})_1 (2p_{3/2}^2)_2]_3$	&	4862.6	&	4861.4	$\pm$	0.1	&	0.42	$\pm$	0.06	&	0.44	\\
		22	&	B	&	DR	&	$[(1s 2s^2 2p_{1/2})_1 (2p_{3/2}^2)_2]_1$	&	4869.7	&	4869.3	$\pm$	0.2	&	-0.17	$\pm$	0.24	&	-0.20	\\
		23	&	B	&	DR	&	$[(1s 2s^2 2p_{1/2})_0 (2p_{3/2}^2)_0]_0$	&	4874.3	&	blend	&	-0.16	$\pm$	0.02	&	-0.23	\\
		&	B	&	DR	&	$[(1s 2s^2 2p_{1/2})_1 (2p_{3/2}^2)_2]_2$	&	4875.7	&				&				&		\\
		&	B	&	DR	&	$[(1s 2s^2 2p_{1/2})_1 (2p_{3/2}^2)_0]_1$	&	4877.4	&				&				&		\\
		24	&	B	&	TR	&	$[1s 2s^2 2p_{3/2}^3]_2$	&	4888.3	&	4887.3	$\pm$	0.1	&	-0.43	$\pm$	0.05	&	-0.45	\\
		25	&	B	&	TR	&	$[1s 2s^2 2p_{3/2}^3]_1$	&	4898.9	&	4897.6	$\pm$	0.1	&	0.15	$\pm$	0.03	&	0.23	\\
		26	&	Be	&	TR	&	$[(1s 2s)_0 2p_{3/2}^3]_{3/2}$	&	4908.6	&	fix.			&	-0.35	$\pm$	0.66	&	-0.68	\\
		27	&	Be	&	TR	&	$[1s 2s^2 (2p_{3/2}^2)_2]_{5/2}$	&	4923.2	&	4919.5	$\pm$	0.4	&	0.41	$\pm$	0.12	&	0.44	\\
		28	&	C	&	DR	&	$[1s 2s^2 2p_{1/2}^2 (2p_{3/2}^2)_2]_{5/2}$	&	4925.7	&	4925.6	$\pm$	0.1	&	0.37	$\pm$	0.05	&	0.43	\\
		29	&	C	&	DR	&	$[1s 2s^2 2p_{1/2}^1 2p_{3/2}^2]_{3/2}$	&	4935.5	&	fix.			&	-0.29	$\pm$	0.52	&	-0.70	\\
		30	&	C	&	DR	&	$[1s 2s^2 2p_{1/2}^2 (2p_{3/2}^2)_0]_{1/2}$	&	4937.7	&	fix.			&	-0.06	$\pm$	0.44	&	0.00	\\
		31	&	C	&	TR	&	$[1s 2s^2 2p_{1/2} (2p_{3/2}^3)_2]_{5/2}$	&	4952.2	&	calib.			&	norm.			&	-0.55	\\
		32	&	C	&	TR	&	$[(1s 2s^2 2p_{1/2})_1 2p_{3/2}^3]_{1/2}$	&	4962.2	&	blend	&	0.18	$\pm$	0.03	&	0.08	\\
		&	C	&	TR	&	$[(1s 2s^2 2p_{1/2})_1 2p_{3/2}^3]_{3/2}$	&	4964.0	&				&				&		\\
		33	&	B	&	TR	&	$[(1s 2s^1 2p_{1/2}^2)_1 (2p_{3/2}^2)_0]_1$	&	4974.0	&	4971.7	$\pm$	1.1	&	0.24	$\pm$	0.17	&	-0.03	\\
		34	&	B	&	QR	&	$[((1s 2s)_0 2p_{1/2})_{1/2} 2p_{3/2}^3]_2$	&	4986.7	&	4984.0	$\pm$	1.0	&	0.03	$\pm$	0.21	&	-0.11	\\
		35	&	N	&	DR	&	$[1s 2s^2 2p_{1/2}^2 2p_{3/2}^3]_2$	&	5010.5	&	5010.2	$\pm$	0.5	&	-0.03	$\pm$	0.13	&	-0.03	\\
		36	&	N	&	DR	&	$[1s 2s^2 2p_{1/2}^2 2p_{3/2}^3]_1$	&	5018.0	&	5017.7	$\pm$	0.8	&	-0.01	$\pm$	0.18	&	-0.04	\\
		37	&	N	&	TR	&	$[(1s 2s^2 2p_{1/2})_1 2p_{3/2}^4]_{1}$	&	5036.7	&	5036.0	$\pm$	1.0	&	0.20	$\pm$	0.21	&	0.12	\\
		38	&	C	&	QR	&	$[1s 2s 2p_{1/2} 2p_{3/2}^4]_{1/2}$	&	5079.0	&	fix.			&	-0.10	$\pm$	0.53	&	-0.20	\\
		39	&	O	&	DR	&	$[1s 2s^2 2p_{1/2}^2 2p_{3/2}^4]_{1/2}$	&	5080.1	&	5081.1	$\pm$	0.7	&	0.01	$\pm$	0.12	&	0.00	\\
		\hline
		\label{tab:results}
	\end{tabular}
\end{table*}

\section{Comparison with theory}\label{sec:theory}
In the following theoretical treatment, we consider DR as a two--step resonant process:
\begin{eqnarray}
\label{eq_DR_description}
\mathrm{e}^- (E_\mathrm{e} l j) &+& \mathrm{A}^{q+}(\alpha_i J_i) \nonumber\\
&& \hspace*{-1cm} \to \mathrm{A}^{(q-1)+ *}(\alpha_d J_d) \to  \mathrm{A}^{(q-1)+}(\alpha_f J_f) + \hbar\omega \, .
\end{eqnarray}
Here, the first step is the resonant capture of a free electron with the energy $E_\mathrm{e}$ and total angular momentum $j$ by an initial ion $\mathrm{A}^{q+}$ in the charge state $q$. This capture leads to the formation of an excited ion $\mathrm{A}^{(q-1)+ *}$ which has a charge state reduced by one unit. In the second step, this excited ion decays to the ground state under emission of characteristic X-rays. In Eq.~(\ref{eq_DR_description}), $J_i$, $J_d$ and $J_f$ represent the total angular momenta of the initial, intermediate excited, and final states, respectively. Additional quantum numbers needed for a full state characterization are collected in variables $\alpha_i$, $\alpha_d$, and $\alpha_f$.

\subsection{Resonance energy and strength}

We computed the electronic structure of each ground and excited state of the ion with FAC v1.1.1 in a fully relativistic way and used the distorted-wave approximation for the interaction with continuum states (see Ref.~\citep{gu2008} and references therein). 
We included the full relativistic form of the electron-electron interaction (Coulomb + Breit interaction) in the atomic potential perceived by the free, incoming (distorted) electron~\citep{breit1929}. 
Higher-order electronic processes cannot be described by the independent particle model; these processes can only be mediated by configuration mixing of the intermediate bound states. 
Thus, to treat them, full-order configuration mixing between the excited states was included in our calculations. 
For the present {KLL} DR calculations, $1s^{2} (2l)^{e}$, $1s (2l)^{e+2}$, and $1s^{2} (2l)^{e+1}$ configurations were included for the initial, intermediate, and final states, respectively. 
From He-like to O-like charge states, the number of electrons $e$ in the L-shell changes from none up to six. 
Furthermore, extended sets of configurations were included in the initial ionic configuration space in order to assign higher-order resonances unidentified in a previous experiment~\citep{beilmann_pa2013}. 
In an extended calculation, configurations with principle quantum numbers up to $n$=5 and all their possible angular momentum $l$-states are included. By that, we take all possible higher excited states such as KLM, KLN, KLO, etc into account. However, we note that it causes negligible changes in the originally predicted values.

The DR cross sections and resonance strengths were calculated in the isolated resonance approximation. 
The DR strength can be written as
\begin{equation}
S^\mathrm{DR} = \frac{\pi^2 \hbar^3 }{m_\mathrm{e} E_{\mathrm{res}}} \frac{g_d}{2g_i} A_a \frac{\sum A_r}{\sum A_r + \sum A_a},
\label{eq:DRstrength}
\end{equation}
where $m_\mathrm{e}$ is the electron mass. $g_i$ and $g_d$ are the statistical weights of the initial and intermediate states. $E_{\mathrm{res}}$ is the resonance energy. $A_a$ represents the autoionization rate from the intermediate excited state to the initial state and $A_r$ represents the radiative rate from the intermediate state to the final state. 

Tables~\ref{tab:he-like}-\ref{tab:o-like} in appendix~\ref{appendix} list all calculated resonances, energies ($E_\mathrm{res}$), and resonance strengths ($S^\mathrm{DR}$) for initially He-like to O-like Fe ions labeled by the order of the recombination processes such as DR, TR, and QR.

For each resonance observed in the present experiment, the extracted resonance energies are compared with the FAC calculations in Tab.~\ref{tab:results}. 
We found an excellent agreement between theory and experiment. For well-resolved resonances, an average deviation of $-1.3$ eV was found between theory and experiment.

\subsection{X-ray anisotropy and polarization}

In the following theoretical treatment we calculate X-ray polarization only for allowed electric dipole $E1$ transitions. 
This channel usually dominates the radiative decay for low-Z few-electron ions where contributions from higher-order multipole components are negligible compared to the $E1$ component.

In DR, the resonant capture of an electron by an ion usually leads to a non-statistical population of magnetic sublevels. 
Thus, the X-rays following resonant capture processes are anisotropic and polarized~\citep{chen1995,beiersdorfer1996}. 
The non-statistical magnetic sublevel population of the intermediate excited states is most naturally described in terms of alignment parameter $\mathcal{A}_2$. 
Following the density matrix formalism described by~\citet{hamiltonbook}, for the resonant capture of an unpolarized electron, the alignment parameter $\mathcal{A}_2$ can be expressed in the form of:
\begin{equation}
\mathcal{A}_{2} (J_d) = \sum_{m} (-1)^{J_d + m} \sqrt{5(2J_d + 1)} \begin{pmatrix}
J_d & J_d & 2\\
-m & m & 0 \end{pmatrix} \sigma_{m}.
\label{eq:alignment}
\end{equation}
Here, the quantity in large parentheses represents the Wigner 3-$j$ symbol, \textit{m} is the magnetic quantum number, and $\sigma_{m}$ are the partial magnetic sublevel cross sections, normalized such that $\sum_{m} \sigma_m = 1$. 
Since both the electron beam and the ions are unpolarized, then $\sigma_{m}=\sigma_{-m}$. 
In this case, as in the present experiment, intermediate excited states with $J_d=0$ or $J_d=1/2$ cannot be aligned and they decay emitting isotropic and unpolarized X-ray. 
Here, the partial magnetic sublevel cross sections $\sigma_m$ were evaluated using the distorted-wave formalism used in FAC, and the relativistic corrections to the electron-electron interaction are accounted for by including the Breit interaction in the zero energy limit for the exchanged virtual photon (cf. FAC v1.1.1 manual and~\citet{gu2008}).

Owing to the alignment of the intermediate excited ion, subsequent DR X-ray emission is polarized and anisotropic. 
Thus, for $E1$ transitions, the degree of polarization and angular distribution are given by~\citep{balashovbook,surzhykov2006}
\begin{equation}
P^\mathrm{DR} (\theta) = - \frac{3 \mathcal{A}_2 \alpha_2^{df} \sin^2 \theta}{2 - \mathcal{A}_2 \alpha_2^{df} (1 - 3 \cos^2 \theta)},
\label{eq:pol} 
\end{equation}
\begin{equation}
I^{DR}(\theta) \propto 1 + \mathcal{A}_2 \alpha_2^{df} P_2(\cos\theta)
\label{eq:ang}
\end{equation}
where $\theta$ denotes the angle of the X-rays with respect to the direction of the electron beam propagation (quantization axis), and $P_2(\cos\theta)$ is the second order Legendre polynomial.  $\alpha_{2}^{df}$ represents an intrinsic anisotropy parameter which is completely determined by the total angular momentum of the intermediate excited state $J_{d}$ and the final state $J_{f}$. 
It can be expressed as~\citep{balashovbook}
\begin{equation}
\alpha_{2}^{df}(J_d, J_f) = (-1)^{J_d + J_f -1} \sqrt{\frac{3(2J_d + 1)}{2}}
\begin{Bmatrix}
1 & 1 & 2\\
J_d & J_d & J_f \end{Bmatrix},
\label{eq:alpha}
\end{equation}
where the quantity in large parenthesis denotes the Wigner 6-$j$ symbol. 
It should be noted that the intermediate excited state in DR usually decays into several close-lying different final states with various degrees of polarization. 
The energy resolution of the Ge-detector used in the present experiments was insufficient to resolve the energy splitting between these final states. Therefore, only the superpositions of individual intermediate to final state transitions are observed. 
Due to this fact, we calculated radiative transition rate weighted ''effective'' intrinsic anisotropy parameters $\bar{\alpha}_2^{df} = \sum_f {\alpha}_2^{df} A_r^{df} / \sum_f A_r^{df}$.

We extracted the emission asymmetries $\mathcal{R}$ by observing X-ray intensities at two different angles with respect to the electron beam propagation axis. 
According to Eqs.~\eqref{eq:P90} and~\eqref{eq:ang}, in the present experiment, we have determined the product $\mathcal{A}_2 \bar{\alpha}_2^{df}$. 
Previously, this was also referred to as anisotropy parameter $\beta$ by~\citet{chen1995,surzhykov2002,weber2010}. 
The product is given in terms of the experimental $\mathcal{R}$ by
\begin{equation}
\mathcal{A}_2 \bar{\alpha}_2^{df} = - \frac{2\mathcal{R}}{3-\mathcal{R}}.
\label{eq:beta}
\end{equation}
The same product $\mathcal{A}_2 \bar{\alpha}_2^{df}$ also defines the X-ray polarization, see Eq.~\eqref{eq:pol}. 
Therefore, within the leading $E1$ approximation the degree of linear polarization of X-rays emitted at the angle $\theta$ with respect to the electron beam propagation axis can be expressed as
\begin{equation}
P(\theta) = \frac{\mathcal{R}\sin^2\theta}{1 + \mathcal{R}\cos^2\theta}.
\label{eq:p2}
\end{equation}
This equation indicates that the linear polarization of characteristic X-rays emitted in the direction perpendicular to the collision axis coincides with the emission asymmetry, i.e.,~\mbox{$P(90^{\circ}) = \mathcal{R}$}. 
Therefore, the degree of linear polarization of the DR X-rays can be obtained easily by using our angle-resolved measurements.

The comparison of experimental values of $\mathcal{R}$ with FAC results is outlined in Table~\ref{tab:results} as well as in the top panel of Fig.~\ref{fig:1Dspec}. 
Note that in the case of an unresolved intermediate excited state (blended resonance), the ''weighted'' emission asymmetry $\mathcal{R}$ is compared with the experimental value. It is calculated according to the following formula:
\begin{equation}
\mathcal{R} = \frac{\sum_i {S_i^{\mathrm{DR}}}{\mathcal{R}_i}}{\sum_i {S_i^{\mathrm{DR}}}}.
\end{equation}
Here, $S^{\mathrm{DR}}$ is the resonance strengths calculated according to Eq.~\ref{eq:DRstrength}. 
{Moreover, we found negligible influence of the Breit interaction on the degree of linear polarization and resonance strengths for Fe ions when compared to the calculations without the Breit interaction.}

All calculated theoretical values of alignment parameters $\mathcal{A}_2$, intrinsic anisotropy parameters ${\alpha}_2^{df}$, and degrees of linear polarization $P^{\mathrm{DR}}$ of X-rays are listed in the Tabs.~\ref{tab:alpha}-\ref{tab:o-like} in appendix~\ref{appendix}.

\subsection{Synthetic spectrum}

Table~\ref{tab:results} compares experiment and theory for each of the individual resonances observed. 
Further, in order to account for all unresolved and weak transitions we compare the experimental spectra with the synthetic one constructed with the help of all theoretical values presented in appendix~\ref{appendix}, see the bottom panels of the Fig.~\ref{fig:1Dspec}. 

The differential resonance strength $dS/d\Omega$ shown in Fig.~\ref{fig:1Dspec} were obtained by multiplying the total resonance strengths with the angular correction factors at respective angles 90$^{\circ}$ and 0$^{\circ}$ and with the charge state distribution present in the experiment. 
By taking the ratio between the observed intensity of a ''well-resolved'' resonance formed by a particular initial ionic state~$I_{\mathrm{CS}}$ to the theoretical strength of that resonance~$S_{\mathrm{CS}}$, we obtained the normalized ion abundance ${n}_{\mathrm{CS}}$ from our experiment. 
For example, in case of He-like initial ions, the normalized ionic abundance can be expressed by:
\begin{equation}
	{n}_{\mathrm{He}} = {\left[\frac{I_{\mathrm{He}}(90^\circ)}{S_{\mathrm{He}}(90^\circ)}\right]}
	\bigg/
	{\sum_{\mathrm{CS}} \left[ \frac{I_{\mathrm{CS}}(90^\circ)}{S_{\mathrm{CS}}(90^\circ)} \right]},
\end{equation}
where $\mathrm{CS}$ refers to charge states from He-like to O-like iron ions. 

According to this equation, for scan (a), we have obtained a distribution with 17\% He-like, 12\% Li-like, and 13\% Be-like Fe ions. 
Likewise, for scan (b), fractional abundances of 20\%, 25\%, 24\%, and 15\% were found for B-, C-, N-, and O-like Fe ions, respectively. 
With this, the charge-state normalized differential resonance strengths $dS/d\Omega$ were obtained for each resonance.
We considered $dS/d\Omega$ as line amplitudes, resonance energies $E_\mathrm{res}$ as line centroids, and then both were convoluted to widths equal to the experimental electron beam energy resolution in order to construct the synthetic spectra (a) and (b). 
The bottom panel of figure~\ref{fig:1Dspec} shows that with some exceptions these calculations indeed very well agree with the present experiment. 
Some caveats with respect to the modeled spectrum can be seen in the part (a) of the spectrum. 
This could be due to the lower collision energy resolution in measurement (a) where many blended resonances introduce errors in reliable extraction of ion abundances. 
The isolated resonance approximation in the theoretical strengths calculations also ignores the possible interference between two resonances. 
We also note that the interference between RR and DR was not included in constructing synthetic spectra. 
However, it is expected to be negligible for low-Z ions. 
%

%
\section{Application of present data: diagnostic of electron cyclotron motion in the hot plasmas}\label{sec:application}
%
%

As discussed before, anisotropic and polarized line radiation can be produced by ions excited by interactions with a directional beam of electrons. 
Thus, the presence of a directional electron component in a laboratory and astrophysical plasma can be confirmed by a polarization measurement. 
Here we use the resonant recombination polarization data to diagnose the electron cyclotron energy component of the electron beam propagating through the magnetic field of the EBIT. 
For the analysis, we combined our present~\ion{Fe}{19}-\ion{}{25} polarization data with the previously reported~\ion{Kr}{29}-\ion{}{35} polarization data by~\citet{shah2016,amaro2017}.

\begin{figure}
	\centering
	\includegraphics[width=0.92\columnwidth]{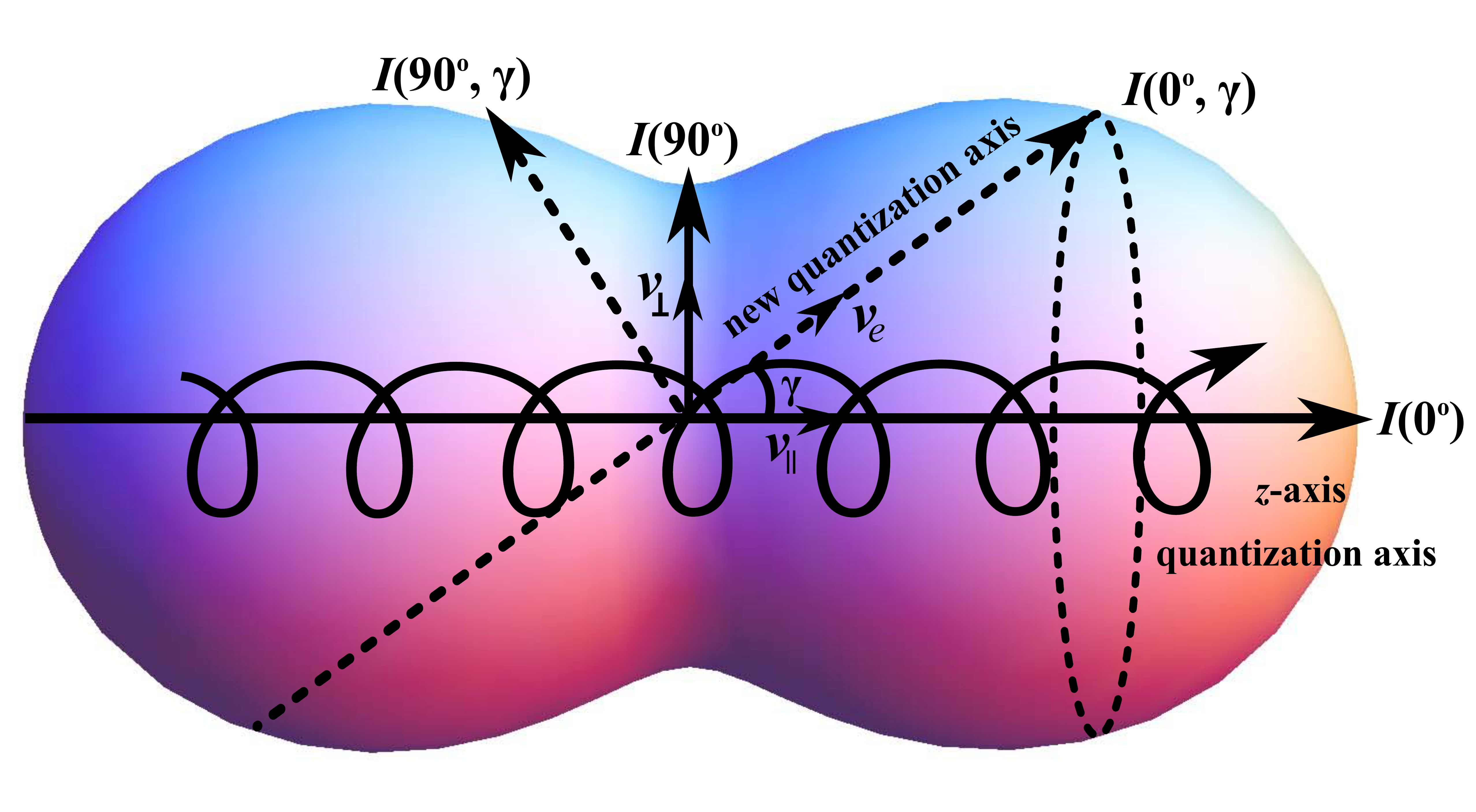}
	\caption{Emission asymmetry of the emitted photons with an introduction of the pitch angle $\gamma$, which is an angle between the magnetic field line (quantization axis) and electron velocity vector $\vec{\nu_{e}}$.}
	\label{fig:pitchangle}
\end{figure}

In the EBIT, the electron beam is not truly laminar and unidirectional since it is produced at the hot cathode. 
The latter emits electrons with a non-vanishing thermal component. 
This causes electrons to gyrate perpendicular to the magnetic field lines and follow helical paths collinear with the electron beam direction. 
This situation may also exist in astrophysical plasmas such as solar flares where electrons spiral around strongly directed magnetic field lines~\citep{haug1972,haug1979,haug1981}. 
The electron beam properties of an EBIT can be described using the non-laminar optical theory of~\citet{herrmann1958}. 
According to this theory, a transverse velocity component of a given electron in motion is related to its radial distance from the cathode center and its velocity at birth on the cathode; and the magnitude of this velocity component is inversely proportional to the radius of cathode images formed at different locations along the beam axis. 
This means the product of transversal energy $E_{\perp}$ and electron beam area is constant along the beam propagation direction, i.e., $E_{\perp} \cdot 2 \pi r^2 \approx$ {constant}~\citep{herrmann1958,beiersdorfer1996,beiersdorfer2001}.
Therefore, by using the cathode temperature ($k_\mathrm{B} T_c$) and the ratio of the beam radius at the cathode ($r_c$) to the beam radius at the trap center ($r_t$) we can estimate $E_{\perp}$ at the trap center.
{
It can be expressed as}
\begin{equation}
	{{E_{\perp}} = {k_\mathrm{B} T_c} \,\,\, \mathrm{[eV]}  \left(\frac {r_c}{r_t}\right)^{2}.}
	\label{eq:E_perp}
\end{equation}
{Because of this transversal component, the instantaneous velocity vector is no longer aligned with the magnetic field or $z$-axis of the EBIT, but it deviates by an angle $\gamma$ from the $z$-axis, as shown in Fig.~\ref{fig:pitchangle}. 
This pitch angle between the $z$- axis and electron velocity vector $\vec{\nu_{e}}$ can be given as
\begin{equation}
	{\gamma  = \mathrm{sin}^{-1} \sqrt{\frac{E_\perp}{E_\mathrm{e}}},}
	\label{eq:sinfunc}
\end{equation}
where $E_{e}$ is the total electron beam energy.}

As shown in Fig.~\ref{fig:pitchangle}, the ions interact with electrons with velocity vectors at an angle $\gamma$ which introduces a new quantization axis to the electron-ion collision system. 
Because of this, X-ray detectors see either an increase or decrease in the total intensity of the emitted X-rays. 
This is the case in the present experiment.
{
A modified expression of Eq.~\eqref{eq:ang} can be derived to characterize this change according to~\citet{gu1999}. 
Considering electrons in an EBIT with a pitch angle distribution $f(\cos\gamma)$ normalized according to \mbox{$\int_{-1}^{1} f(\cos\gamma) \,\, d(\cos\gamma) = 1$}, the resulting angular distribution of DR X-rays produced by spiraling beam of electrons can be written as
\begin{equation}
{I^{\mathrm{DR}}(\theta, \gamma) \propto 1 + \mathcal{C}_2 \mathcal{A}_{2} \alpha_{2}^{df} P_2(\cos\theta),}
\label{eq:ang-pitch}
\end{equation}
where 
\begin{equation}
{\mathcal{C}_2 = \int_{-1}^{1} f(\cos\gamma) \,\, d(\cos\gamma) \,\, P_2(\cos\gamma).}
\label{eq:c2}
\end{equation}
For a typical pitch angle $\gamma_0$ in an EBIT, the pitch angle distribution can be described by a Dirac delta function, i.e.,~\mbox{$f(\cos\gamma) = \delta(\cos\gamma - \cos\gamma_0)$}~\citep{gu1999}. In this case, according to Eq.~\eqref{eq:ang-pitch} and Eq.~\eqref{eq:c2}, the observed emission asymmetry $\mathcal{R}$ of DR X-rays can be expressed as
\begin{equation}
{\mathcal{R} = \frac{I(90^\circ, \gamma)-I(0^\circ,\gamma)}{I(90^\circ, \gamma)}
			= \frac{3 \mathcal{A}_{2} \bar\alpha_{2}^{df} (1 - 3 \cos^2\gamma)}{4 + \mathcal{A}_{2} \bar\alpha_{2}^{df} (1 - 3 \cos^2\gamma)}.}
			\label{eq:pitchnotzero}
\end{equation}
}

{
In the ideal case, we would assume that the electron beam propagating inside an EBIT is perfectly unidirectional and the pitch angle $\gamma$ is zero ($\gamma = 0$). The emission asymmetry $\mathcal{R}_0$ would then be expressed as
\begin{equation}
	\mathcal{R}_0 = \frac{I(90\,^{\circ}) - I(0\,^{\circ})}{I(90\,^{\circ})} = -\frac{3\mathcal{A}_{2} \bar\alpha_{2}^{df}}{2- \mathcal{A}_{2} \bar\alpha_{2}^{df}},
	\label{eq:pitchzero}
\end{equation}
where \mbox{$f(\cos\gamma) = \delta(\cos\gamma - 1)$}, i.e., according to Eq.~\eqref{eq:c2}, \mbox{$\mathcal{C}_2 = 1$}.
Under the actual conditions, by combining equations~\eqref{eq:pitchzero} and \eqref{eq:pitchnotzero}, the \textit{real} emission asymmetry $\mathcal{R}$ takes the form of
\begin{equation}
	\mathcal{R} = \frac{\mathcal{R}_0 (3\sin^2\gamma-2)}{\mathcal{R}_0 \sin^2\gamma - 2}.
\end{equation}
}

{
We now compare the values of $\mathcal{R}$ determined from the experiment (where $\gamma$ is not zero) to the theoretical $\mathcal{R}_0$ value for which $\gamma$ was assumed to be zero. 
Thereby, we determine the experimental value of transversal energy component $E_\perp$ and corresponding pitch angle $\gamma$ of the electron beam:}
\begin{equation}
{\gamma = {\sin}^{-1} \sqrt{\frac{2(\mathcal{R} - \mathcal{R}_0)}{\mathcal{R}_0(\mathcal{R}-3)}},}
\label{eq:pitchexp}
\end{equation}

\begin{equation}
{E_{\perp} = \frac{2(\mathcal{R} - \mathcal{R}_0)}{\mathcal{R}_0(\mathcal{R}-3)} \,\,\, E_\mathrm{e}.}
\label{eq:perpexp}
\end{equation}

\begin{figure}
	\centering
	\includegraphics[width=0.92\columnwidth]{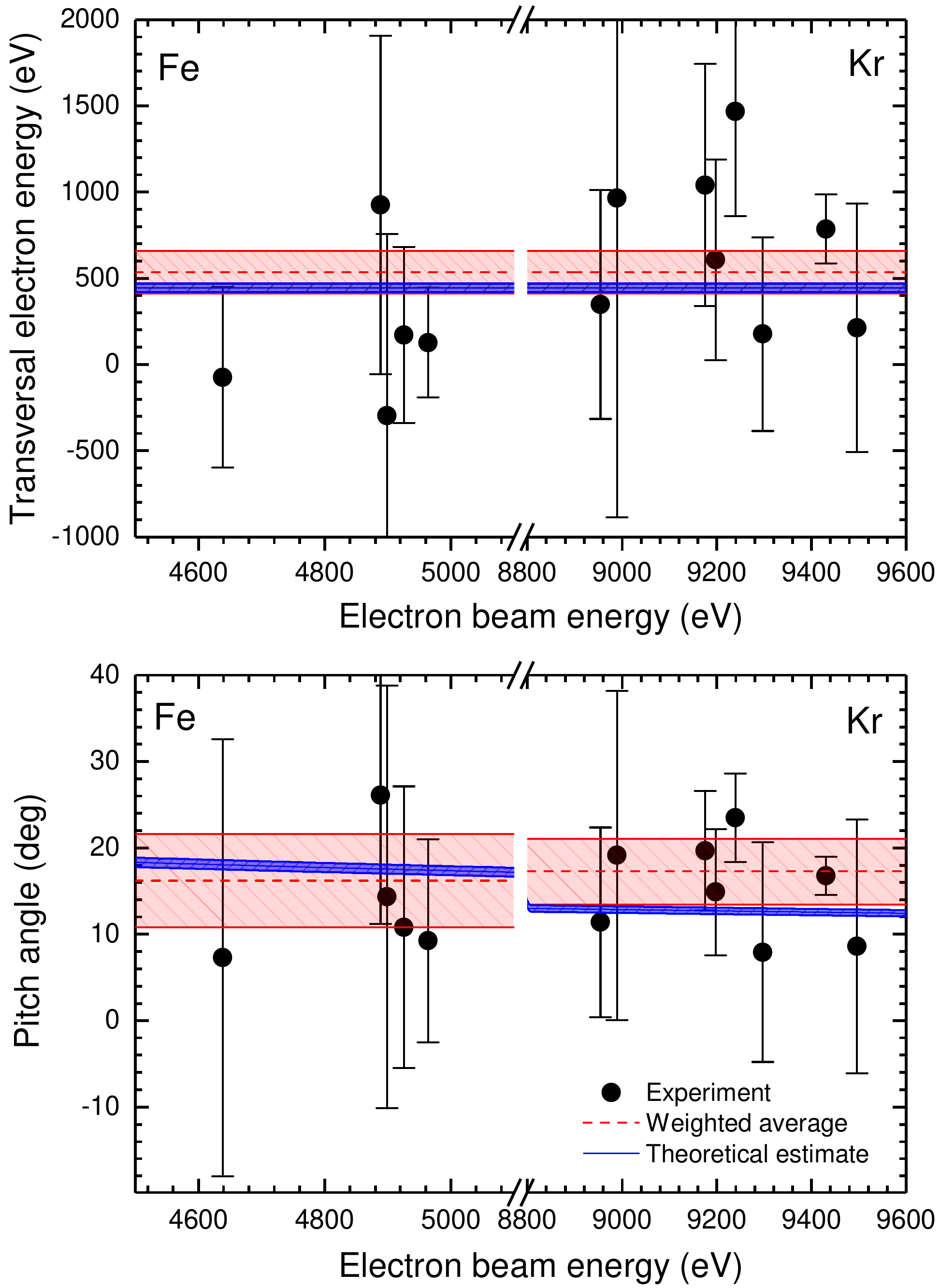}
	\caption{Experimental $E_\perp$ and corresponding $\gamma$ as a function of the electron beam energy (see Eqs.~\eqref{eq:pitchexp} and~\eqref{eq:perpexp}). The red shaded area represents the weighted average value and containing uncertainty. The blue shaded area represents the theory estimated according to the experimental electron beam conditions: cathode radius $r_c$ = 1.5 mm, magnetic field $B$ = 6 T, residual magnetic field at cathode $B_c$ = 300--500 $\mu$T, and cathode temperature $T_c$ = 1200--1400 K.}
	\label{fig:EXP_Eperp_vs_KLL}
\end{figure}

For each well-resolved (not blended) resonance, we determined values of $E_\perp$ and $\gamma$ by applying above Eqs.~\eqref{eq:pitchexp} and~\eqref{eq:perpexp}. 
The experimental uncertainties are computed as a quadrature sum of errors associated with quantities in these equations. 
The extracted values of $E_\perp$ and $\gamma$ from each resonance are graphically shown in the Fig.~\ref{fig:EXP_Eperp_vs_KLL}. 
According to~\eqref{eq:sinfunc}, the transversal energy component of the electron beam does not change drastically over a certain electron beam energy range. 
{
For a complete scan range of both Fe and Kr measurements ($E_\mathrm{e}=$ 4--10 keV), we give a single weighted average value of $E_\perp$ which is found to be $534 \pm 124$ eV.
For the pitch angle $\gamma$, we present separate weighted average values for both experiments because it changes over the scan range of electron beam energies (see Eq.~\eqref{eq:pitchexp}). 
In case of the electron beam energy of 4.5--5.1 keV in the Fe measurement, we obtained the average value of $\gamma = 16.3^\circ \pm 5.4^\circ$, whereas in the energy range of 8.8--9.6 keV, used in the Kr measurement, we obtained $\gamma = 17.2^\circ \pm 2.8^\circ$. 
The weighted average values are shown as red shaded area in the Fig.~\ref{fig:EXP_Eperp_vs_KLL}. 
}

The theoretical values of $E_\perp$ and $\gamma$ can be estimated based on the electron beam conditions. 
{The cathode with a diameter of 3 mm is heated to approximately 1200--1400 K, resulting in a transverse thermal energy component of $\approx$ 0.10--0.12 eV at the cathode. }
As mentioned before, the electron beam of the FLASH-EBIT is compressed by the magnetic field with a strength of 6 T. 
During the measurements, we positioned the electron gun and used correction coil surrounding the electron gun to ensure nearly zero residual magnetic field at the cathode. 
{However, this residual magnetic field is not very well known. Based on the observed beam diameter, which is close to the Herrmann value for negligible residual magnetic field, we estimate an upper bound of the order a few hundred microteslas (300--500 $\mu\mathrm{T}$ assumed here)}. 
With these assumptions, theoretical values of $E_\perp$ and $\gamma$ were estimated according to Eqs.~\eqref{eq:E_perp} and~\eqref{eq:sinfunc}. 
They are shown as blue shaded area in the Fig.~\ref{fig:EXP_Eperp_vs_KLL}. 
It can be seen that the theoretical estimates according to Hermann's optical theory are in good accord with the experimentally determined average values of $E_\perp$ and $\gamma$. 
{Furthermore, the electron beam radius $r_t$ at the trap was also deduced using the experimental value of $E_\perp$ and Eq.~\ref{eq:E_perp}. 
The radius was found to be $22.6 \pm 4.2~\micron$ which is also in a very good agreement with the Herrmann's theory, see Tab.~\ref{tab:pars_fe}. }
This validation reassures our measurements and estimates since the optical theory of Herrmann was used as one of the first principles in the design of the electron beam ion trap~\citep{levine1988}. 
For comparison, we also estimated the transversal energy component of the electron beam based on the principle of adiabatic magnetic flux invariance, i.e., $\nu_\perp^2/B \approx$ constant. 
According to this, $E_\perp$ was estimated to be 1578 eV. Such large value of transversal energy does not agree with our measured value. 
This can be explained since the EBIT has a non-uniform magnetic field along the electron beam propagation axis. 
The electrons travel across strong magnetic field gradients from few microteslas magnetic field at the electron gun to 6~teslas at the trap.
Such gradients invalidate the applicability of the principle of adiabatic magnetic flux invariance~\citep{beiersdorfer2001}.
%

\section{Summary and Conclusions}\label{sec:summary}

In this work, we have performed comprehensive experimental and theoretical studies on the polarization of X-rays produced by {KLL} DR of highly charged He-like to O-like iron ions. 
The emitted radiation was recorded along and perpendicular to the electron beam propagation axis and the X-ray polarization of dielectronic recombination satellite lines were extracted from the measured emission asymmetries.
The presented experimental technique greatly reduces the time required for comprehensive measurements compared to direct polarization measurements which are time-consuming and require an X-ray polarimeter~\citep{weber2015,shah2015}. 
Our method is simple to implement at EBITs and it allows to scan all ions relevant to astrophysical and fusion research. It can be applied not only to the DR process, but also to measure X-ray polarization due to other atomic processes. 
With a few exceptions, we found excellent agreement between the experiment and distorted-wave predictions of FAC. 
We thus show that this code is reliable to a high degree, and can be used to produce large sets of atomic data and to develop reliable astrophysical plasma models. 
A contribution of ubiquitous, but hitherto unrecognized, higher-order channels, such as TR, was found to be as strong as the dielectronic process in B- and C-like Fe ions, and it should play an important role in the charge balance determination of hot plasmas~\citep{beilmann2011,beiersdorfer2015}. 
Recently, TR channels have also proven to play a dominant role in the measurable X-ray polarization of a common type of astrophysical and laboratory plasmas~\citep{shah2016}.

In addition, we have demonstrated the suitability of resonant recombination polarization measurements for diagnostics of anisotropies of hot plasmas, such as plasmas of solar flares~\citep{haug1972,haug1979}.
In our demonstration we probed the electron cyclotron motion component of the electron beam propagating through a strong magnetic field of the FLASH-EBIT. 
The electron cyclotron energy was accounted for $534 \pm 124$ eV of total electron beam energy. Furthermore, the electron beam radius at the trap center was determined to be $22.6 \pm 4.2~\micron$. Both values agree exceptionally well with the predictions of the optical theory of electron beams by~\citet{herrmann1958}. 
Such agreement shows that Hermann theory can be useful to predict the systematic effect of the electron cyclotron motion component on polarization measurements.
This analysis is far more complex and extracts even greater details of the plasma anisotropies than the one which would be required to detect the directional component of the hot plasmas of a solar flare or a tokamak fusion plasma. Therefore, the present work opens numerous possibilities for diagnostics of the anisotropies of hot laboratory and astrophysical plasmas. 

With the next generation of X-ray satellites, namely~\textit{XARM} (X-ray Astronomy Recovery Mission) and \textit{Athena}, exceptional spectral energy resolutions will be reached with higher sensitivity~\citep{kelley2016,hitomi2016}. 
The present data will be particularly useful for interpreting high-resolution microcalorimeter based X-ray spectra in the future~\citep{hitomi2017}. 
It can also be used to benchmark spectral modeling codes for hot collisional plasmas such as SPEX~\citep{kaastra1996}, CHIANTI~\citep{delzanna2015}, and AtomDB~\citep{foster2012}. 
Furthermore, these results may also be applicable to model X-ray line polarization data from the future X-ray polarimetry missions such as ESA M4 class mission -- \textit{XIPE}~\citep{soffitta2013}, NASA Small Explorer mission -- \textit{PRAXyS}~\citep{jahoda2014}, and \textit{IXPE}~\citep{weisskopf2016}. 
For example, the photoelectric gas polarimeter~\citep{costa2001} of \textit{XIPE} will not resolve individual X-ray line transitions. 
Such limitations are rather general even when the high-resolution detector will be utilized because Doppler shifts will most likely obscure the signal. 
In such instances, the line polarization will contain contributions from many transitions induced by different atomic processes. 
This calls for more quantitatively reliable experimental and theoretical atomic data of X-ray polarization of all contributing channels.

\section{Acknowledgments}
\acknowledgments
This work was supported by the Deutsche Forschungsgemeinschaft (DFG) within the Emmy Noether program under Contract No. TA 740 1-1 and by the Bundesministerium f\"ur Bildung und Forschung (BMBF) under Contract No. 05K13VH2. 



\newpage
\clearpage

\appendix

\section{KLL DR calculations of \ion{Fe}{19}-\ion{}{25}}\label{appendix}

Here, we present theoretical DR data that are tested and benchmarked within the present experiment. 
The following quantities are presented in the Tables~\ref{tab:alpha}--\ref{tab:o-like}. 
First column of Tab.~\ref{tab:he-like}--\ref{tab:o-like} represents energy level index. 
Second column identifies the order of recombination processes, DR: Dielectronic Recombination, TR: Trielectronic Recombination, QR: Quadruelectronic Recombination. 
An intermediate excited state configuration of the ion is given in third column. Standard \textit{j-j} coupling notations are used here to define atomic configurations. The subscript following the round bracket denotes angular momentum of the coupled subshells and another subscript following the square bracket denotes the total angular momentum of the given state. 
Resonance energies $E_{\mathrm{res}}$ in eV and total resonance strengths $S^{\mathrm{DR}}$ in units of $\mathrm{cm}^{2}$~eV are given in fourth and fifth columns, respectively. The former value is the energy difference between the initial and intermediate excited ionic states, and the latter values are calculated using equations~\eqref{eq:DRstrength}. 
To calculate degree of linear polarization, the alignment parameter $\mathcal{A}_2$ is given in sixth column with taking frequency-independent Breit interaction into account, and intrinsic anisotropy parameter $\alpha_{2}^{df}$ for the X-ray transition from intermediate excited state to final states is given in Tab.~\ref{tab:alpha}.  
As discussed in Sec.~\ref{sec:theory}, intermediate excited state produced by DR usually decays to several final states. 
Final state configuration are given in the seventh column, accordingly, the X-ray energies $E$\textsubscript{X-ray} and radiative decay rates $A_r^{df}$ (in s$^{-1}$) are provided in next two columns, respectively. 
Finally, the polarization of X-rays $P^\mathrm{DR}$ is calculated according to equation~\eqref{eq:pol} and is given in the last column. 
Note that for the experimental comparison, we used "effective" intrinsic anisotropy parameter $\bar \alpha_{2}^{df}$ which is weighted by given radiative rates $A_r^{df}$.

\begin{table}[!h]
	\centering
	\setlength{\tabcolsep}{1.5em}
	\caption{Intrinsic anisotropy parameter $\alpha_2^{df}$ (Eq.~\eqref{eq:alpha}) characterizing X-ray polarization and anisotropy.}
\begin{tabular}{ccc|ccc}
		\hline
		$J_d$ & $J_f$ & $\alpha_2^{df}$ & $J_d$ & $J_f$ & $\alpha_2^{df}$ \\
		\hline
		0     & 1     & 0     & $1/2$ & $1/2$ & 0 \\
		1     & 0     & $1/\sqrt{2}$ & $1/2$ & $3/2$ & 0 \\
		1     & 1     & $-1/2\sqrt{2}$ & $3/2$ & $1/2$ & $1/2$ \\
		1     & 2     & $1/10\sqrt{2}$ & $3/2$ & $3/2$ & $-2/5$ \\
		2     & 1     & $\sqrt{7}/2\sqrt{10}$ & $3/2$ & $5/2$ & $1/10$ \\
		2     & 2     & $-\sqrt{7}/2\sqrt{10}$ & $5/2$ & $3/2$ & $\sqrt{7}/5\sqrt{2}$ \\
		2     & 3     & $1/\sqrt{70}$ & $5/2$ & $5/2$ & $-4\sqrt{2}/5\sqrt{7}$ \\
		3     & 2     & $\sqrt{3}/5$ & $5/2$ & $7/2$ & $1/2\sqrt{14}$ \\
		3     & 3     & $-\sqrt{3}/4$ &       &       &  \\
		3     & 4     & $-1/4\sqrt{3}$ &       &       &  \\
		\hline
	\end{tabular}%
	\label{tab:alpha}%
\end{table}%

\begin{longtable}{@{\extracolsep\fill}cclccclccc}
		\caption{\ion{Fe}{25} theoretical data. Initial (ground) state: $[1s^2]_0$}\\
		\hline
		$i$   & Process & Intermediate excited & $E$\textsubscript{res} & $S^{\mathrm{DR}}$ & $\mathcal{A}_{2}$ & Final & $E$\textsubscript{X-ray} & $A_r^{df}$ & $P^{\mathrm{DR}}$ \\
		&       & state & (eV)  & ($\mathrm{cm}^2 \mathrm{eV}$) &       & state & (eV)  &  (s$^{-1}$) &  \\
		\hline
		\endfirsthead
		\multicolumn{10}{@{}c}{\tablename{~\ref{tab:he-like}}---\textit{Continued}}\\
		\hline
		$i$   & Process & Intermediate excited & $E$\textsubscript{res} & $S^{\mathrm{DR}}$ & $\mathcal{A}_{2}$ & Final & $E$\textsubscript{X-ray} & $A_r^{df}$ & $P^{\mathrm{DR}}$ \\
		&       & state & (eV)  & ($\mathrm{cm}^2 \mathrm{eV}$) &       & state & (eV)  &  (s$^{-1}$) &  \\
		\hline
		\endhead
		\endfoot
		\endlastfoot
			4     & DR    & $[1s (2s^2)_{0}]_{1/2}$ & 4554.41 & 1.982E-20 & 0.000 & $[1s^2 2p_{1/2}]_{1/2}$ & 6552.00 & 9.95E+12 & 0.000 \\
			&       &       &       &       &       & $[1s^2 2p_{3/2}]_{3/2}$ & 6536.14 & 1.05E+13 & 0.000 \\
			5     & DR    & $[(1s 2s)_{1} 2p_{1/2}]_{1/2}$ & 4565.87 & 6.439E-23 & 0.000 & $[1s^2 2s]_{1/2}$ & 6611.86 & 4.96E+12 & 0.000 \\
			6     & DR    & $[(1s 2s)_{1} 2p_{1/2}]_{3/2}$ & 4569.59 & 3.885E-22 & -1.000 & $[1s^2 2s]_{1/2}$ & 6615.57 & 1.59E+13 & 0.600 \\
			8     & DR    & $[(1s 2s)_{0} 2p_{1/2}]_{1/2}$ & 4607.76 & 5.469E-20 & 0.000 & $[1s^2 2s]_{1/2}$ & 6653.75 & 3.38E+14 & 0.000 \\
			9     & DR    & $[(1s 2s)_{1} 2p_{3/2}]_{3/2}$ & 4616.73 & 8.229E-22 & -1.000 & $[1s^2 2s]_{1/2}$ & 6662.72 & 5.04E+14 & 0.600 \\
			10    & DR    & $[1s (2p_{1/2}^2)_{0}]_{1/2}$ & 4624.16 & 8.286E-23 & 0.000 & $[1s^2 2p_{1/2}]_{1/2}$ & 6621.75 & 2.05E+13 & 0.000 \\
			&       &       &       &       &       & $[1s^2 2p_{3/2}]_{3/2}$ & 6605.89 & 1.20E+11 & 0.000 \\
\end{longtable}%

\begin{longtable}{@{\extracolsep\fill}cclccclccc}
	\caption{\ion{Fe}{24} theoretical data. Initial (ground) state: $[1s^2 2s]_{1/2}$ }\\
	\hline
	$i$   & Process & Intermediate excited & $E$\textsubscript{res} & $S^{\mathrm{DR}}$ & $\mathcal{A}_{2}$ & Final & $E$\textsubscript{X-ray} & $A_r^{df}$ & $P^{\mathrm{DR}}$ \\
	&       & state & (eV)  & ($\mathrm{cm}^2 \mathrm{eV}$) &       & state & (eV)  &  (s$^{-1}$) &  \\
	\hline
	\endfirsthead
	\multicolumn{10}{@{}c}{\tablename{~\ref{tab:li-like}}---\textit{Continued}}\\
	\hline
	$i$   & Process & Intermediate excited & $E$\textsubscript{res} & $S^{\mathrm{DR}}$ & $\mathcal{A}_{2}$ & Final & $E$\textsubscript{X-ray} & $A_r^{df}$ & $P^{\mathrm{DR}}$ \\
	&       & state & (eV)  & ($\mathrm{cm}^2 \mathrm{eV}$) &       & state & (eV)  &  (s$^{-1}$) &  \\
	\hline
	\endhead
	\endfoot
	\endlastfoot
		11    & DR    & $[1s 2s^2 2p_{1/2}]_{0}$ & 4644.94 & 7.965E-22 & 0.000 & $[1s^2 2p_{1/2} 2p_{3/2}]_1$ & 6469.08 & 8.57E+12 & 0.000 \\
		12    & DR    & $[1s 2s^2 2p_{1/2}]_{1}$ & 4646.84 & 1.136E-20 & 0.694 & $[1s^2 2s^2]_0$ & 6597.86 & 3.76E+13 & -0.975 \\
		&       &       &       &       &       & $[1s^2 2p_{1/2}^2]_0$ & 6479.63 & 3.90E+12 & -0.975 \\
		&       &       &       &       &       & $[1s^2 2p_{1/2} 2p_{3/2}]_1$ & 6470.98 & 2.15E+12 & 0.328 \\
		&       &       &       &       &       & $[1s^2 2p_{1/2} 2p_{3/2}]_2$ & 6465.17 & 3.90E+12 & -0.075 \\
		13    & DR    & $[1s 2s^2 2p_{3/2}]_{2}$ & 4660.00 & 4.946E-21 & -0.839 & $[1s^2 2p_{1/2} 2p_{3/2}]_1$ & 6484.14 & 3.46E+12 & 0.448 \\
		&       &       &       &       &       & $[1s^2 2p_{1/2} 2p_{3/2}]_2$ & 6478.33 & 6.44E+12 & -0.639 \\
		&       &       &       &       &       & $[1s^2 2p_{3/2}^2]_2$ & 6461.25 & 1.14E+12 & -0.639 \\
\end{longtable}%

\begin{longtable}{@{\extracolsep\fill}cclccclccc}
	\caption{\ion{Fe}{23} theoretical data. Initial (ground) state:  $[1s^2 2s]_{1/2}$ }\\
	\hline
	$i$   & Process & Intermediate excited & $E$\textsubscript{res} & $S^{\mathrm{DR}}$ & $\mathcal{A}_{2}$ & Final & $E$\textsubscript{X-ray} & $A_r^{df}$ & $P^{\mathrm{DR}}$ \\
	&       & state & (eV)  & ($\mathrm{cm}^2 \mathrm{eV}$) &       & state & (eV)  &  (s$^{-1}$) &  \\
	\hline
	\endfirsthead
	\multicolumn{10}{@{}c}{\tablename{~\ref{tab:be-like}}---\textit{Continued}}\\
	\hline
	$i$   & Process & Intermediate excited & $E$\textsubscript{res} & $S^{\mathrm{DR}}$ & $\mathcal{A}_{2}$ & Final & $E$\textsubscript{X-ray} & $A_r^{df}$ & $P^{\mathrm{DR}}$ \\
	&       & state & (eV)  & ($\mathrm{cm}^2 \mathrm{eV}$) &       & state & (eV)  &  (s$^{-1}$) &  \\
	\hline
	\endhead
	\endfoot
	\endlastfoot
		16    & DR    & $[1s 2s^2 2p_{1/2}^2]_{1/2}$ & 4755.78 & 9.320E-22 & 0.000 & $[1s^2 2s^2 2p_{1/2}]_{1/2}$ & 6552.85 & 3.07E+13 & 0.000 \\
		&       &       &       &       &       & $[1s^2 2s^2 2p_{3/2}]_{3/2}$ & 6538.27 & 1.79E+12 & 0.000 \\
		&       &       &       &       &       & $[1s^2 2p_{1/2}^2 2p_{3/2}]_{3/2}$ & 6398.50 & 3.36E+12 & 0.000 \\
		&       &       &       &       &       & $[1s^2 2p_{1/2} 2p_{3/2}^2]_{3/2}$ & 6379.27 & 4.28E+10 & 0.000 \\
		&       &       &       &       &       & $[1s^2 2p_{1/2} 2p_{3/2}^2]_{1/2}$ & 6357.04 & 1.62E+10 & 0.000 \\
		&       &       &       &       &       & $[1s^2 2p_{3/2}^3]_{3/2}$ & 6350.22 & 1.81E+10 & 0.000 \\
		17    & DR    & $[(1s 2s^2 2p_{1/2})_{0} 2p_{3/2}]_{3/2}$ & 4764.68 & 3.164E-23 & -1.000 & $[1s^2 2s^2 2p_{1/2}]_{1/2}$ & 6561.75 & 4.18E+10 & 0.600 \\
		&       &       &       &       &       & $[1s^2 2s^2 2p_{3/2}]_{3/2}$ & 6547.17 & 8.98E+12 & -0.750 \\
\end{longtable}%

\begin{longtable}{@{\extracolsep\fill}cclccclccc}
	\caption{\ion{Fe}{22} theoretical data. Initial (ground) state:  $[1s^2 2s^2 2p_{1/2}]_{1/2}$ }\\
	\hline
	$i$   & Process & Intermediate excited & $E$\textsubscript{res} & $S^{\mathrm{DR}}$ & $\mathcal{A}_{2}$ & Final & $E$\textsubscript{X-ray} & $A_r^{df}$ & $P^{\mathrm{DR}}$ \\
	&       & state & (eV)  & ($\mathrm{cm}^2 \mathrm{eV}$) &       & state & (eV)  &  (s$^{-1}$) &  \\
	\hline
	\endfirsthead
	\multicolumn{10}{@{}c}{\tablename{~\ref{tab:b-like}}---\textit{Continued}}\\
	\hline
	$i$   & Process & Intermediate excited & $E$\textsubscript{res} & $S^{\mathrm{DR}}$ & $\mathcal{A}_{2}$ & Final & $E$\textsubscript{X-ray} & $A_r^{df}$ & $P^{\mathrm{DR}}$ \\
	&       & state & (eV)  & ($\mathrm{cm}^2 \mathrm{eV}$) &       & state & (eV)  &  (s$^{-1}$) &  \\
	\hline
	\endhead
	\endfoot
	\endlastfoot
		21    & DR    & $[1s 2s^2 2p_{1/2}^2 2p_{3/2}]_{2}$ & 4830.72 & 5.975E-22 & -0.897 & $[1s^2 2s^2 2p_{1/2} 2p_{3/2}]_{1}$ & 6509.07 & 9.96E+12 & 0.474 \\
		&       &       &       &       &       & $[1s^2 2s^2 2p_{1/2} 2p_{3/2}]_{2}$ & 6503.59 & 9.31E+12 & -0.693 \\
		&       &       &       &       &       & $[1s^2 2s^2 2p_{3/2}^2]_{2}$ & 6487.56 & 3.27E+11 & -0.693 \\
		&       &       &       &       &       & $[1s^2 2p_{1/2}^2 2p_{3/2}^2]_{2}$ & 6313.38 & 1.32E+11 & -0.693 \\
		&       &       &       &       &       & $[1s^2 2p_{1/2} 2p_{3/2}^3]_{1}$ & 6301.89 & 5.52E+09 & 0.474 \\
		22    & DR    & $[1s 2s^2 2p_{1/2}^2 2p_{3/2}]_{1}$ & 4856.94 & 4.736E-20 & -0.684 & $[1s^2 2s^2 2p_{1/2}^2]_{0}$ & 6544.47 & 2.85E+14 & 0.584 \\
		&       &       &       &       &       & $[1s^2 2s^2 2p_{1/2} 2p_{3/2}]_{1}$ & 6535.29 & 2.63E+12 & -0.412 \\
		&       &       &       &       &       & $[1s^2 2s^2 2p_{1/2} 2p_{3/2}]_{2}$ & 6529.81 & 5.00E+13 & 0.071 \\
\end{longtable}%

\begin{longtable}{@{\extracolsep\fill}cclccclccc}
	\caption{\ion{Fe}{21} theoretical data. Initial (ground) state:  $[1s^2 2s^2 2p_{1/2}^2]_0$ }\\
	\hline
	$i$   & Process & Intermediate excited & $E$\textsubscript{res} & $S^{\mathrm{DR}}$ & $\mathcal{A}_{2}$ & Final & $E$\textsubscript{X-ray} & $A_r^{df}$ & $P^{\mathrm{DR}}$ \\
	&       & state & (eV)  & ($\mathrm{cm}^2 \mathrm{eV}$) &       & state & (eV)  &  (s$^{-1}$) &  \\
	\hline
	\endfirsthead
	\multicolumn{10}{@{}c}{\tablename{~\ref{tab:c-like}}---\textit{Continued}}\\
	\hline
$i$   & Process & Intermediate excited & $E$\textsubscript{res} & $S^{\mathrm{DR}}$ & $\mathcal{A}_{2}$ & Final & $E$\textsubscript{X-ray} & $A_r^{df}$ & $P^{\mathrm{DR}}$ \\
&       & state & (eV)  & ($\mathrm{cm}^2 \mathrm{eV}$) &       & state & (eV)  &  (s$^{-1}$) &  \\
	\hline
	\endhead
	\endfoot
	\endlastfoot
		16    & DR    & $[1s 2s^2 2p_{1/2}^2 (2p_{3/2}^2)_2]_{5/2}$ & 4925.71 & 3.800E-20 & -1.069 & $[1s^2 2s^2 2p_{1/2}^2 2p_{3/2}]_{3/2}$ & 6497.29 & 2.04E+14 & 0.500 \\
		&       &       &       &       &       & $[1s^2 2s^2 2p_{1/2} 2p_{3/2}^2]_{3/2}$ & 6479.47 & 1.48E+13 & 0.500 \\
		&       &       &       &       &       & $[1s^2 2s^2 2p_{1/2} 2p_{3/2}^2]_{5/2}$ & 6474.72 & 1.76E+13 & -0.889 \\
		&       &       &       &       &       & $[1s^2 2s^2 2p_{3/2}^3]_{3/2}$ & 6456.86 & 1.25E+11 & 0.500 \\
		&       &       &       &       &       & $[1s^2 2p_{1/2}^2 2p_{3/2}^3]_{3/2}$ & 6252.29 & 7.30E+10 & 0.500 \\
		17    & DR    & $[(1s 2s^2 2p_{1/2})_{0} (2p_{3/2}^3)_{3/2}]_{3/2}$ & 4935.53 & 9.324E-21 & -1.000 & $[1s^2 2s^2 2p_{1/2}^2 2p_{3/2}]_{3/2}$ & 6507.10 & 2.42E+14 & -0.750 \\
		&       &       &       &       &       & $[1s^2 2s^2 2p_{1/2} 2p_{3/2}^2]_{3/2}$ & 6489.29 & 6.87E+12 & -0.750 \\
		&       &       &       &       &       & $[1s^2 2s^2 2p_{1/2} 2p_{3/2}^2]_{5/2}$ & 6484.54 & 1.19E+13 & 0.143 \\
\end{longtable}%

\begin{longtable}{@{\extracolsep\fill}cclccclccc}
	\caption{\ion{Fe}{20} theoretical data. Initial (ground) state:  $[1s^2 2s^2 2p_{1/2}^2 2p_{3/2}]_{3/2}$ }\\
	\hline
	$i$   & Process & Intermediate excited & $E$\textsubscript{res} & $S^{\mathrm{DR}}$ & $\mathcal{A}_{2}$ & Final & $E$\textsubscript{X-ray} & $A_r^{df}$ & $P^{\mathrm{DR}}$ \\
	&       & state & (eV)  & ($\mathrm{cm}^2 \mathrm{eV}$) &       & state & (eV)  &  (s$^{-1}$) &  \\
	\hline
	\endfirsthead
	\multicolumn{10}{@{}c}{\tablename{~\ref{tab:n-like}}---\textit{Continued}}\\
	\hline
	$i$   & Process & Intermediate excited & $E$\textsubscript{res} & $S^{\mathrm{DR}}$ & $\mathcal{A}_{2}$ & Final & $E$\textsubscript{X-ray} & $A_r^{df}$ & $P^{\mathrm{DR}}$ \\
	&       & state & (eV)  & ($\mathrm{cm}^2 \mathrm{eV}$) &       & state & (eV)  &  (s$^{-1}$) &  \\
	\hline
	\endhead
	\endfoot
	\endlastfoot
		11    & DR    & $[1s 2s^2 2p_{1/2}^2 2p_{3/2}^3]_{2}$ & 5010.46 & 3.597E-21 & -0.100 & $[1s^2 2s^2 2p_{1/2}^2 2p_{3/2}^2]_{2}$ & 6466.76 & 2.97E+14 & 0.348 \\
		&       &       &       &       &       & $[1s^2 2s^2 2p_{1/2} 2p_{3/2}^3]_{1}$ & 6455.73 & 1.09E+14 & -0.453 \\
		&       &       &       &       &       & $[1s^2 2s^2 2p_{1/2} 2p_{3/2}^3]_{2}$ & 6445.45 & 3.02E+13 & 0.348 \\
		12    & DR    & $[1s 2s^2 2p_{1/2}^2 2p_{3/2}^3]_{1}$ & 5017.98 & 1.721E-21 & 0.143 & $[1s^2 2s^2 2p_{1/2}^2 2p_{3/2}^2]_{2}$ & 6474.28 & 2.26E+14 & -0.059 \\
		&       &       &       &       &       & $[1s^2 2s^2 2p_{1/2}^2 2p_{3/2}^2]_{0}$ & 6465.02 & 1.44E+14 & -0.723 \\
		&       &       &       &       &       & $[1s^2 2s^2 2p_{1/2} 2p_{3/2}^3]_{1}$ & 6463.25 & 9.35E+13 & 0.265 \\
		&       &       &       &       &       & $[1s^2 2s^2 2p_{1/2} 2p_{3/2}^3]_{2}$ & 6452.97 & 3.66E+13 & -0.059 \\
		&       &       &       &       &       & $[1s^2 2s^2 2p_{3/2}^4]_{0}$ & 6434.16 & 8.80E+11 & -0.723 \\
\end{longtable}%

\begin{longtable}{@{\extracolsep\fill}cclccclccc}
	\caption{\ion{Fe}{19} theoretical data. Initial (ground) state:  $[1s^2 2s^2 2p_{1/2}^2 2p_{3/2}^2]_{2}$ }\\
	\hline
	$i$   & Process & Intermediate excited & $E$\textsubscript{res} & $S^{\mathrm{DR}}$ & $\mathcal{A}_{2}$ & Final & $E$\textsubscript{X-ray} & $A_r^{df}$ & $P^{\mathrm{DR}}$ \\
	&       & state & (eV)  & ($\mathrm{cm}^2 \mathrm{eV}$) &       & state & (eV)  &  (s$^{-1}$) &  \\
	\hline
	\endfirsthead
	\multicolumn{10}{@{}c}{\tablename{~\ref{tab:o-like}}---\textit{Continued}}\\
	\hline
	$i$   & Process & Intermediate excited & $E$\textsubscript{res} & $S^{\mathrm{DR}}$ & $\mathcal{A}_{2}$ & Final & $E$\textsubscript{X-ray} & $A_r^{df}$ & $P^{\mathrm{DR}}$ \\
	&       & state & (eV)  & ($\mathrm{cm}^2 \mathrm{eV}$) &       & state & (eV)  &  (s$^{-1}$) &  \\
	\hline
	\endhead
	\endfoot
	\endlastfoot
	13 & DR    & $[1s 2s^2 2p_{1/2}^2 2p_{3/2}^4]_{1/2}$ & 5080.14 & 3.968E-21 & 0.000 & $[1s^2 2s^2 2p_{1/2}^2 2p_{3/2}^3]_{3/2}$ & 6434.99 & 4.26E+14 & 0.000 \\
		&       &       &       &       &       & $[1s^2 2s^2 2p_{1/2} 2p_{3/2}^4]_{1/2}$ & 6422.33 & 2.13E+14 & 0.000 \\
	\hline
	\label{tab:o-like}%
\end{longtable}%

\end{document}